\documentclass[twocolumn,aps,pra,amssymb,amsfonts,letterpaper,nofootinbib]{revtex4}
\def\dist{{\rm dist}}
\usepackage{hyperref}
\usepackage{enumerate}
\usepackage{verbatim}
\usepackage{amsmath}
\usepackage{latexsym}
\usepackage{revsymb}
\usepackage{yfonts}
\usepackage{amsfonts}
\usepackage{amsmath}
\usepackage{amssymb}
\usepackage{amsthm}
\usepackage{graphicx}
\usepackage{bm}
\usepackage{bbm}
\usepackage{color}
\usepackage{mathrsfs}
\usepackage{epstopdf}

\usepackage{color}

\def\copies{l}

\def\duzomniejsze{<\kern-.7mm<}
\def\duzowieksze{>\kern-.7mm>}

\def\textbf#1{{\bf #1}}
\def\beq{\begin{equation}}
\def\eeq{\end{equation}}
\def\be{\begin{equation}}
\def\ee{\end{equation}}
\def\ben{\begin{eqnarray}}
\def\een{\end{eqnarray}}
\def\beqa{\begin{eqnarray}}
\def\eeqa{\end{eqnarray}}
\def\eea{\end{array}}
\def\bea{\begin{array}}
\newcommand{\bei}{\begin{itemize}}
\newcommand{\eei}{\end{itemize}}
\newcommand{\bee}{\begin{enumerate}}
\newcommand{\eee}{\end{enumerate}}

\def\hcal{{\cal H}}

\def\ccal{{\mathbb C}} %cal

\def\tr{{\rm Tr}}

\def\>{\rangle}
\def\<{\langle}

\def\ot{\otimes}

\newtheorem{lemma}{Lemma}

\newtheorem{proposition}{Proposition}
\newtheorem{theorem}{Theorem}
\newtheorem{definition}{Definition}

\def\bed{\begin{definition}}
\def\eed{\end{definition}}
\def\bel{\begin{lemma}}
\def\eel{\end{lemma}}
\def\bet{\begin{theorem}}
\def\eet{\end{theorem}}

\begin{document}

 \title{Bound entangled states with extremal properties}

\begin{abstract}
Following recent work of Beigi and Shor, we investigate  PPT states that are ``heavily entangled.''  
We first exploit volumetric methods to show that in a randomly chosen direction, 
there
 are PPT states
whose distance in trace norm from separable states is  
(asymptotically) at least $1/4$. 
We then provide explicit examples of PPT states which are nearly as far from
separable ones as possible. To obtain
% distance 
a distance of
$2-\epsilon$ \ 
 from
 the
  separable states, we need 
  %dimension
  a dimension of
    $2^{{\rm poly}(\log({1\over \epsilon}))}$, 
as opposed to $2^{{\rm poly}({1\over \epsilon})}$ given by the construction  
of Beigi and Shor \cite{Beigi-Shor}. We do so by exploiting the so called {\it private states}, 
 introduced earlier in the context of quantum cryptography. 
We also provide 
a
lower bound for the distance between private states 
and PPT states and investigate the distance between
pure states and the set of PPT states.
\end{abstract}
\author{Piotr Badzi\c{a}g$^{(1)}$, Karol Horodecki$^{(2)}$, Micha\l{} Horodecki$^{(3)}$, Justin Jenkinson$^{(4)}$ and Stanis\l{}aw J. Szarek$^{(4,5)}$ }

\affiliation{$^{(1)}$Physics Department, Stockholm University, S-0691 Stockholm, Sweden}
\affiliation{$^{(2)}$Institute of Informatics,
University of Gda\'nsk, 80--952 Gda\'nsk, Poland}
\affiliation{$^{(3)}$Institute of Theoretical Physics and Astrophysics, 
University of Gda\'nsk, 80--952 Gda\'nsk, Poland}
\affiliation{$^{(4)}$Case Western Reserve University, Cleveland, Ohio 44106-7058, USA}
\affiliation{$^{(5)}$Universit\'e Pierre et Marie Curie-Paris 6, 75252 Paris, France}

\maketitle 
 
The set of PPT states (i.e. states with positive partial transpose) plays an 
important role in quantum information theory. 
While the PPT criterion perfectly discovers entanglement in pure states and for 
$2\otimes 2$ and  $2\otimes 3$ systems, it is not always conclusive \cite{rmp-ent}  
in higher dimensions. 
The entangled states that have the PPT property are known to be {\it bound entangled} :  no pure entanglement  can be distilled from them. It is a longstanding open problem whether this last property is equivalent to PPT (see \cite{nptbound-few-steps} and references therein).  
On the other hand, it is possible to obtain cryptographic key from 
some PPT states \cite{pptkey}. In view of such operational characteristics -- 
or conjectured characteristics -- of the  set of PPT states, 
it was often used as a first approximation of the set of separable states. 

The geometric properties of sets of PPT states ($\mathcal{PPT}$) and that 
of separable states ($\mathcal{SEP}$) were investigated
starting with \cite{ZyczkowskiHSP-vol,GB1,SBZ} 
Recently there has been interest in quantifying 
how different $\mathcal{PPT}$ 
and $\mathcal{SEP}$ are.   It was 
shown in \cite{AubrunSzarek}  that the ratio of the volumes of 
$\mathcal{PPT}$ and $\mathcal{SEP}$ 
grows super-exponentially in the dimension of the sets. 
The distance between a PPT state and $\mathcal{SEP}$ 
%the set of separable states 
was investigated in \cite{Beigi-Shor}, 
where it was proved that there exist PPT states that lie as far from separable 
states as it is possible, namely $2-\epsilon$ 
in {\em trace norm distance},\footnote{Here and further in this paper by trace norm distance we mean $\|\rho-\sigma\|_1$ 
were $\|\cdot\|_1$ is the trace norm. In \cite{Beigi-Shor} the distance with 
an additional factor $1/2$ was used.} for any positive $\epsilon$, 
provided the dimension is large enough. 
Thresholds for the PPT property and for separability for random induced states 
were compared in \cite{ASY} (see also \cite{Aubrun, ASY1}) and shown to be 
dramatically different.

In this paper we will revisit the phenomena studied in  \cite{Beigi-Shor}. 
First, we will show how similar  results can be deduced by well-known methods 
from the values of various geometric invariants of $\mathcal{PPT}$ and $\mathcal{SEP}$ 
calculated in  \cite{AubrunSzarek}. A sample result states that a ``generic witness'' 
can detect a PPT state whose separability violation is about  $1/4$. 
%(on the scale  corresponding to the trace distance considered in  \cite{Beigi-Shor}. 
Next, we provide an alternate (explicit) construction of a family of states 
that recovers the $2-\epsilon$ bound from  \cite{Beigi-Shor} 
and show how their dimensions scale depending on $\epsilon$.  

With regards to the construction, our argument is based on {\em private states}, which  
were introduced in order to investigate the relationship between quantum security
and entanglement \cite{pptkey}. They have been already used in 
the context of cryptography \cite{RenesSmith06,LiWinterZouGuo}, as well as in channel theory \cite{Smith-Yard,GeneralUncSec,GeneralUncSecIEEE}. Here
we use this class to investigate the geometry of the set $\mathcal{PPT}$.
The general idea is that {\em every} private state $\gamma$ is ``rather 
far''  from any separable state \cite{karol-PhD,keyhuge}.
If we can show that some PPT state $\rho$ is not ``too far'' from $\gamma$, 
we obtain easily a lower 
bound on
the 
 distance between $\rho$ and the set $\mathcal{SEP}$ of separable states. 
Similarly as in \cite{Beigi-Shor}, our construction involves 
taking tensor power of some chosen state (here
it is the one constructed in  \cite{smallkey}). 
However, we do not use  tools such as de Finetti theorem or quantum tomography, 
but instead rely on simple permanence properties of the sets in question.
Our construction is essentially self-contained;  
it vastly  improves the scaling of the dimension needed to obtain 
distance $2-\epsilon$, which  
in our case is $2^{C (\log{\frac{4}{\epsilon}})^2}$, with $C<6$. 
(Here and in what follows all logarithms are to the base of $2$.) 
No explicit formula  is given in \cite{Beigi-Shor}, but 
an examination of the argument presented there shows that 
it requires  the dimension to be of order $2^{({1/\epsilon})^\kappa}$,
where $\kappa$ is at least $2$  (and probably larger).

We also analyze limitations of  the approach via private states due to the fact that, 
in finite dimension, there is always a nonzero gap between private states and PPT states.
We obtain a lower bound on this gap in the 
case of $\ccal^{2d}\ot\ccal^{2d}$ states (private bits), extending results of \cite{KimSanders}.
We also find that the ``distance'' of  pure states from PPT states  in terms of {\em fidelity} 
 is the same as that from separable states.
This shows that our construction could not work  with the set of pure states 
instead of the set of private states  as a starting point.

%In the paper 
In this work we use the following notation. 
For a state $\rho_{AB}$ on a composite system 
$A \otimes B$ we denote the {\em partial transpose} on system
 $B$ as  $\rho_{AB}^\Gamma:=({\rm Id}\ot T)( \rho_{AB})$,  
 where $T$ is the transpose map  
 (while the result does depend on what system we perform 
 the partial tranpose, its positivity does not). 
 We denote the {\em trace norm} by $\|X\|_1:=\tr \sqrt{XX^\dagger}$ and, 
 more generally, the {\em $p$-Schatten norm} by  $\|X\|_p:=\big(\tr (XX^{\dagger})^{p/2}\big)^{1/p}$. 
When talking about  the distance of a state $\rho$ 
from the set of separable states we will always mean the quantity 
\be
\dist(\rho,\mathcal{SEP}):= \min_{\sigma \in \mathcal{SEP}} \|\rho -\sigma\|_1\, . 
\ee
However, analogous expressions for other norms, and for properties 
diffrent from PPT and separability,  may be of interest and can also 
be studied by some of the methods we employ below.

\section{Bounds based on global geometric invariants} 

One of the results of  \cite{AubrunSzarek} (Theorem 1) asserts that for $\ccal^{d}\ot\ccal^{d}$, 
the ratio of the volumes of $\mathcal{PPT}$ and $\mathcal{SEP}$ is at least 
$(cd)^{m/2}$, where $m:=d^4-1$ is the dimension of these sets and $c>0$ is 
a universal (explicit and not too small) constant.  This implies immediately that there 
is a  PPT state $\rho$ whose {\em robustness} (\cite{VidalT}) is at least of order $d^{1/2}$: 
if $\epsilon > c^{-1}d^{-1/2}$, then the mixture $\epsilon \rho + (1-\epsilon){\bf{I}}/d^2$ 
is entangled. (The same assertion holds 
with the maximally mixed state ${\bf{I}}/d^2$ 
replaced by any other separable state $\sigma$, which is called in \cite{VidalT} 
robustness {\em relative to} $\sigma$.) 

\begin{figure}[h]
\begin{center}  
%\includegraphics[height=5cm,angle=0]{intuition_2D.jpg}  
%\hspace{1cm}
\includegraphics[width=9cm,angle=0]{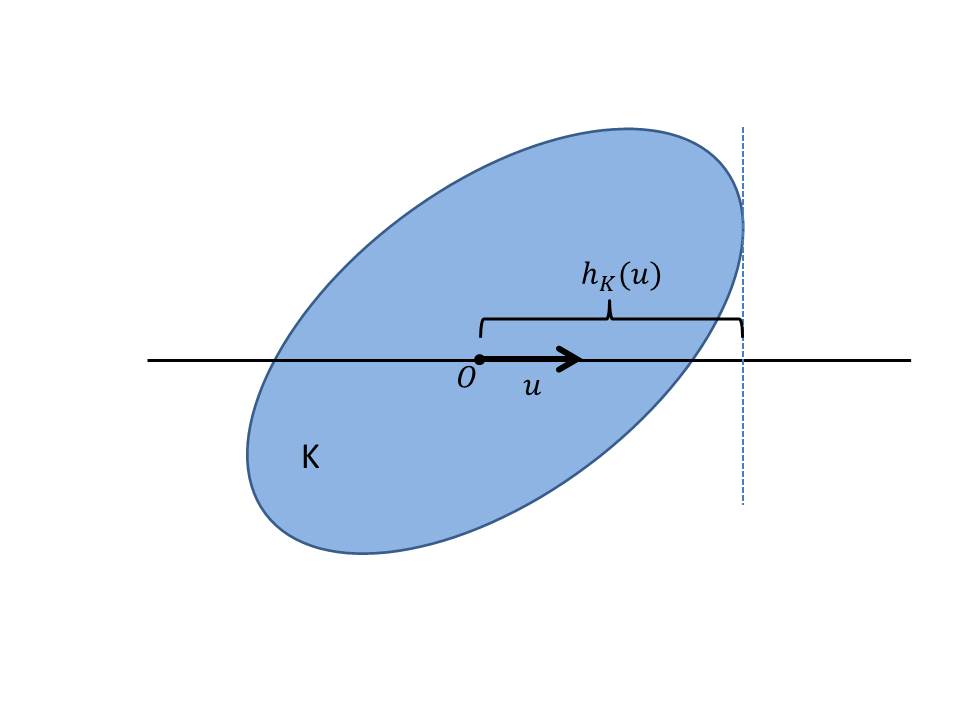}  
\caption{The geometric meaning of  $h_K(u)$, the width of $K$ in 
the direction of $u$. The function $h_K$ is often called 
	``the support function of $K$.'' 
\label{fig:hk}}  
\end{center}  
\end{figure}

The geometric invariant that played more fundamental role than volume 
in the arguments  of  \cite{AubrunSzarek} was the {\em mean width}, 
which is defined as follows. If $K$ is a subset of a (real) Euclidean space 
we define its mean width (actually mean half-width), denoted $w(K)$, as 
\beq \label{def:width}
w(K):= \int_S \max_{x \in K} \< x,u\>\, du \, ,
\eeq
where $S$ is the unit sphere of the space in question and the integration 
is performed with respect to the normalized invariant measure on $S$.  
For a given unit vector $u \in S$, the expression
\beq \label{hK}
h_K(u) := \max_{x \in K} \< x,u\>
\eeq
is usually called the width of $K$ in the direction of $u$ 
(``the {\em extent} of $K$ in the direction of $u$'' would be perhaps 
more appropriate). See Fig. \ref{fig:hk} for graphical interpretation of the quantity.

The mean width is related to the volume by the classical inequality of Urysohn 
${\rm vrad}(K) \leq w(K)$, where ${\rm vrad}(K)$ (the volume radius of $K$) 
is the radius of a Euclidean ball whose volume is equal to that of $K$. 

The asymptotic order of the mean widths of $\mathcal{PPT}$ and 
$\mathcal{SEP}$ -- with respect to the Euclidean structure induced by the 
Hilbert-Schmidt (or Frobenius) norm and 
 as the dimension goes to infinity in various regimes -- 
was determined in \cite{AubrunSzarek}.  For the bipartite systems 
$\ccal^{d}\ot\ccal^{d}$ we have the inequalities (valid for all $d$) 
\beq \label{wsep}
\frac 16 d^{-3/2} \leq {\rm vrad}(\mathcal{SEP})\leq w(\mathcal{SEP}) \leq 4 d^{-3/2} , 
\eeq
\beq \label{wppt}
\frac 14 d^{-1} \leq {\rm vrad}(\mathcal{PPT}) \leq w(\mathcal{PPT}) \leq 2 d^{-1} ,
\eeq
and the limit relation
\beq \label{wpptlim}
\liminf d \,w(\mathcal{PPT}) \geq \liminf d \, {\rm vrad}(\mathcal{PPT})  \geq \frac 1 2 .
\eeq
The details of some of the calculations 
that lead to the specific numerical values 
of the multiplicative constants that appear above are contained in 
\cite{Szarek,SWZ,Justin-PhD}. In fact, the expectation is that 
the limits $\lim d \,w(\mathcal{PPT})$ and   $\lim d^{3/2} \,w(\mathcal{SEP})$ 
exist (and, {\em a posteriori}, belong to the intervals $[\frac 12, 2]$ 
and $[\frac 16, 4]$ respectively), but we do not know of a rigorous 
argument to that effect. By comparison, the precise asymptotic order 
of the mean width of the set of {\em all} states on $\ccal^{n}$ is 
{\em known} to be $2 n^{-1/2}$ 
(i.e., $2d^{-1}$ in our setting; 
that's where the upper estimate in (\ref{wppt}) comes from). 
However, even this fact far from being obvious: the reason for the factor $2$  
is the ``radius $2$'' in Wigner's Semicircle Law \cite{Wig1,Wig2}; 
cf. Lemma \ref{wigner} below and the comments following it.  

As it turns out, much more information is available in addition to   
the bounds on the averages of the width functions  of 
$h_{\mathcal{PPT}}$ and $h_{\mathcal{SEP}}$ given 
by (\ref{wsep})-(\ref{wpptlim}):  one has essentially the same 
{\em pointwise} estimates for $h_{\mathcal{PPT}}(u)$ and 
$h_{\mathcal{SEP}}(u)$ for all but a very small fraction 
of directions $u\in S$.  This is a consequence of the 
classical Levy's {\em  concentration inequality}.  
{\lemma (\cite{Levy, MS86}, \cite{Justin-PhD}) \label{levy} 
Let $m>2$ and let $f$ be an $L$-Lipschitz
function on the sphere $S$ in the $m$-dimensional Euclidean space. 
Then, for every $t>0$,
\beq \label{levy:conc}
P(|f-M| >t ) \leq   \exp(-\frac{m t ^2}{2L^2}),
\eeq
 where $M$ is  the median of $f$ and $P$ is the normalized 
 invariant measure on $S$. 
}

\smallskip For functions of the form (\ref{hK}), the Lipschitz constant $L$ equals 
the outradius of $K$. The outradius of the set of {\em all} states 
on  $\ccal^{n}$ is $\sqrt{1-\frac 1n} < 1$ (provided the center 
of the circumscribed sphere is chosen to be at  the maximally mixed state ${\bf{I}}/d^2$, 
which is the natural choice) and so -- for width functions of sets of states such as 
$f=h_{\mathcal{PPT}}$ or 
$f=h_{\mathcal{SEP}}$ --  the constant $L$ disappears 
from the estimate. Since the dimension of the space 
is  then $m=d^4-1$, it follows that   the probability in (\ref{levy:conc})
is  small  if $t\gg d^{-2}$.  Still another elementary consequence of 
(\ref{levy:conc}) is that the mean and the median of $f$ differ   
at most by $O(m^{-1/2}L)= O(d^{-2})$, 
and so  we can conclude that, for any $\alpha  >0$, 
\ben \label{width:conc}
P\big(|h_{\mathcal{PPT}} -  w({\mathcal{PPT}})| > \alpha d^{-1}\big)  &\leq&  2 \exp(-c\alpha^2d^2)\nonumber \\
P\big(|h_{\mathcal{SEP}} -  w({\mathcal{SEP}})| > \alpha d^{-3/2}\big) & \leq&  2 \exp(-c\alpha^2d), 
%\nonumber 
\een
where $c>0$ is an (explicit) universal constant.  
In particular, if $d$ is sufficiently large, then with probability close to $1$,
\beq \label{extent}
h_{\mathcal{PPT}}(u) >( \frac12 -\alpha)d^{-1},\  h_{\mathcal{SEP}}(u) < (4 +\alpha)d^{-3/2}. 
\eeq
In other words, for  large $d$, the width (or extent) of $\mathcal{PPT}$ 
in most of directions is (at least) about  $\big(\frac12-o(1)\big) d^{-1}$. 
This may not seem very impressive, but should be compared 
with the asymptotic value of $2d^{-1}$ that we obtain by the 
same argument for  the set of {\em all} states. 
On the other hand, the extent of $\mathcal{SEP}$ in a typical direction is of order 
$d^{-3/2} \ll d^{-1}$. 

For our final observation in the spirit of the results of \cite{Beigi-Shor} 
we need another version of Wigner's Semicircle Law, very closely related 
to the asymptotic expression  $2 n^{-1/2}$ for the mean width of the set of 
all states on $\ccal^{n}$ 
{\lemma \label{wigner} Let $S$ be the unit sphere in the space of traceless    
$n \times n$ Hermitian matrices. Then $\int_S \|u\|_\infty \, du$ is 
asymptotically of order $2n^{-1/2}$. Moreover, for any $\epsilon > 0$, 
\be
P\big(\big|  \|u\|_\infty -  2n^{-1/2}\big| > \epsilon n^{-1/2}  \big) <  2 \exp(-c\epsilon^2n) . 
\ee
}

The estimate on probability follows from the first statement  
and from Lemma \ref{levy} (cf. \cite{DavidsonSzarek1,DavidsonSzarek2}). 
In turn, the first statement 
follows immediately from the well-known facts that 

\bei
\item the expected value 
of the norm of GUE matrices is approximately $2\sqrt{n}$ (see, e.g.,  \cite{AGZ} 
and its references) %and 
\item for 
$1$-homogeneous functionals on an $m$-dimensional space, the ratio 
between the spherical average and the mean with respect to the 
standard Gaussian measure is an explicit factor (depending only on $m$), 
which is approximately $m^{-1/2}$.
\eei

Since $n=d^2$ and $m=n^2-1=d^4-1$, 
 the spherical average of $ \|u\|_\infty$ is 
approximately $2\sqrt{n} \times m^{-1/2} = 
2 d (d^4-1)^{-1/2}$, hence approximately $2d^{-1}$. 
There is a minor issue  
related to the fact that the usual GUE ensemble is defined without the trace $0$ 
restriction, but it can be easily handled.   
See also Appendix F in \cite{Szarek} for an argument showing
that  $2 n^{-1/2}$ is an upper bound for all $n$ 
-- and not just an asymptotic approximation -- and for a discussion 
of error terms.  

\smallskip With this preparation, we are ready to show

{\theorem 
Let $\epsilon > 0$. Then, for $d$ large enough (depending on $\epsilon$),
\be
\max_{\rho\in {\mathcal{PPT}}} \dist(\rho,\mathcal{SEP})  \ge  \frac{1}{4} 
-\epsilon\, . 
\ee
Moreover,  this distance is witnessed in most directions $u \in S$.  
}

\smallskip \noindent 
For the proof, consider any direction $u\in S$ for which 
$h_{\mathcal{PPT}}(u) - h_{\mathcal{SEP}}(u) >  \big(\frac12 -\frac {\epsilon}4 \big)d^{-1}$; 
by (\ref{extent}) and the comments following it this happens with probability  
close to $1$ if $d$ is large. Assume also that $u$ does not belong to 
the (small) exceptional set given by the condition 
from Lemma \ref{wigner}  so that  in particular $ \|u\|_\infty <  (2+\epsilon)n^{-1/2}= (2+\epsilon)d^{-1}$.  Let $\rho \in \mathcal{PPT}$ be such that $\<\rho, u\> = h_{\mathcal{PPT}}(u)$ 
and let $\sigma \in \mathcal{SEP}$ be arbitrary. Then 
\ben
\< \rho - \sigma, u\> &=& \< \rho, u\> - \<  \sigma, u\> \nonumber\\
&=& h_{\mathcal{PPT}}(u)- \<  \sigma, u\> \nonumber\\
&\geq& h_{\mathcal{PPT}}(u)- h_{\mathcal{SEP}}(u)  \nonumber\\
&\geq& \big(\frac12 -\frac {\epsilon}4 \big)d^{-1} . %\nonumber
\een
On the other hand, 
\be 
\< \rho - \sigma, u\> \leq \|\rho - \sigma\|_1 \|u\|_\infty  \leq (2+\epsilon) d^{-1}  \|\rho - \sigma\|_1 .
\ee
Combining these inequalities leads to 
\be
 \|\rho - \sigma\|_1 \geq  \big(\frac12 -\frac {\epsilon}4 \big)/(2+\epsilon)>\frac{1}{4} 
-\epsilon .
\ee
This means that the distance of $\rho$ to $\mathcal{SEP}$ in trace distance 
is at least $\frac{1}{4}  -\epsilon$  and that such distance 
 %and the set  $\mathcal{SEP}$ 
can be certified by nearly all  witnesses $u \in S$ (for  
an appropriate $\rho \in \mathcal{PPT}$, depending on $u$) .

\section{PPT states distant from separable states : a construction based on private states}
In the preceding section we showed that, in sufficiently large dimension, 
 PPT states that are quite far from  the set of separable states are ubiquitous. 
 However, our argument was of a probabilistic nature, hence non-constructive. 
  
In the present section we will give an explicit construction of PPT states that 
are nearly as far from separable states as possible.  
The main result is stated in Theorem \ref{prop:main}, 
which provides a bipartite PPT state whose distance from 
the set of separable states  
is larger than  $2-\epsilon$,  with the dimension of the system 
scaling like $2^{O(\log^2(1/\epsilon))}$, i.e., involving 
the number of qubits that  is polylogarythmic in $1/\epsilon$.  
We thus recover the main result of  \cite{Beigi-Shor}, with a much better dependence 
of $\epsilon$ on the dimension. We also consider limitations of our approach, 
generalizing results of \cite{KimSanders} in Proposition \ref{thm:PPTPS}, which investigates distance between PPT states and the so-called ``private states,''
introduced  originally in the context of quantum cryptography in \cite{pptkey}. 

\subsection{Private, separable and PPT states}
\begin{figure}[h]
\begin{center}  
%\includegraphics[height=5cm,angle=0]{intuition_2D.jpg}  
%\hspace{1cm}
\includegraphics[width=7cm,angle=0]{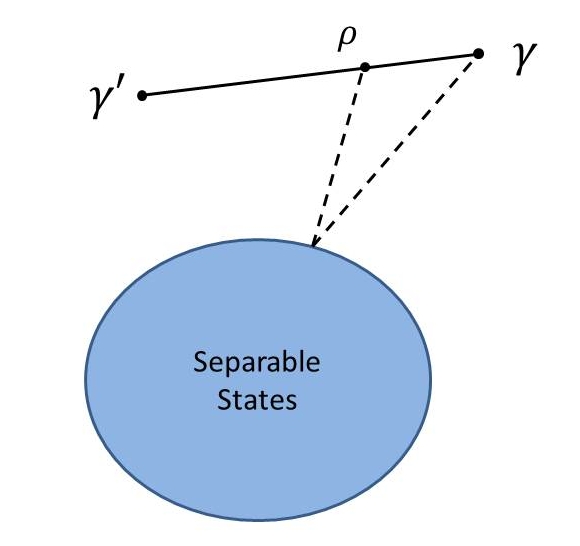}  
\caption{Bounding $\rho$ away from separable states: we show that the 
private state $\gamma$ is far from 
$\mathcal{SEP}$, and that $\rho$ is close to $\gamma$, 
since it  is a mixture  $\gamma$ and $\gamma'$, 
with small weight at $\gamma'$.
\label{fig:dist}}  
\end{center}  
\end{figure}

In our construction of PPT states which are far 
from separable states, we will employ  ``private states'' \cite{pptkey}. 
Their precise definition will be given later in Appendix   \ref{sec:private_states}, 
but here we will only need their features listed below:
\bei
\item any private state is far from separable states, the distance increasing with the dimension 
(Lemma \ref{lem:ps_sep}),
\item at the expense of dimension, there are private states arbitrary close to PPT states 
(Eq. (\ref{eq:close})), 
\item tensor product of private states is again a private state. 
\eei
The idea is now to exploit the first two features and the triangle 
inequality to obtain PPT states whose distance 
to any separable state is at least about $1$  (see Fig. \ref{fig:dist}). 
We next consider tensor powers of that state, which of course remain PPT, 
and show -- by combining the first and the third feature -- that their distance 
to separable states can be boosted as closely to $2$ as desired, 
at the expense of increasing the dimension. 

Private states are states of four systems $A,B,A'$ and $B'$. The systems $A$ and $B$ constitute the {\it key} part, 
while $A'$ and $B'$  - the so-called  {\it shield} part. 
The corresponding Hilbert spaces are $\hcal_A=\hcal_B=C^{d_k}$ and 
%$\hcal_A=\hcal_B=C^{d_s}$
$\hcal_{A'}=\hcal_{B'}=C^{d_s}$.
When $d_k=2$, we will call a private state 
a {\it private bit}.  
%We will not need here explicit definition of private states, 
%but for completeness we provide it in Appendix \ref{sec:private_states}. 
It is immediately seen from the definition  
that the tensor product of two private states is again a
private state, with the key and shield dimensions of the product state being  
products of those of original states.
The following lemma \cite{karol-PhD} (cf. \cite{keyhuge}) 
quantifies the distance of an arbitrary private state 
from the set of separable states.

{\lemma
For any private state $\gamma$ with the key part of 
dimension $d_k\times d_k$ we have 
\be
%\inf_{\sigma \in SEP}\,\|\gamma - \sigma\| 
\dist(\gamma, \mathcal{SEP}) \geq 2 - {2\over d_k}.
\ee
\label{lem:ps_sep}
}
Now set $d_k=2$ and consider  the following  state constructed in \cite{smallkey}
\be
\rho = (1-p) \gamma + p \gamma' ,
\label{eq:state}
\ee
where $p=\frac{1}{\sqrt{d_s}+1}$ and where $\gamma,\gamma'$ are mutually orthogonal private states   given by 
Eq. \eqref{eq:gammas} (Appendix \ref{sec:private_states}). 
The matrix form of $\rho$ is also presented in Eq. \eqref{eq:state2}.
The state $\rho$ has the following properties: (i) it is PPT, since by construction it is invariant under the partial transpose; (ii) it is close to 
the private state $\gamma$  
since  we have 
\be
\|\rho - \gamma\|_1 = 2p = {2 \over \sqrt{d_s}+1}.
\label{eq:close}
\ee
Consider now the closest, in trace norm, separable state to $\rho$, call it $\sigma$. 
Using Lemma \ref{lem:ps_sep} for $d_k=2$ and the triangle inequality we obtain 
\ben
&&\|\rho - \sigma\|_1 + \|\rho - \gamma\|_1 \geq \|\sigma - \gamma\|_1 \geq \nonumber \\ 
&&\geq \dist(\gamma, \mathcal{SEP}) \geq 1
\label{eq:far}
\een
Applying now (\ref{eq:close})	we obtain  the following 
\begin{proposition}  \label{almost1}
Let $\rho$ be the state  given by (\ref{eq:state}). Then its  distance to the 
set of  separable states satisfies 
\be
\dist(\rho,\mathcal{SEP}) \geq 1 -{2 \over \sqrt{d_s}+1}
\label{eq:lower}
\ee
\end{proposition}
We see that this lower bound improves with a larger shield part, 
and is the worst for $d_s = 2$ (then $\rho$ is four-qubit state). In that case we have $\dist(\rho,\mathcal{SEP}) \geq 0.58579$.

It is known that the state $\rho$ lies on the boundary of PPT states 
(see \cite{smallkey} Observation 2), so its choice is in a sense
optimal. To see to what extent the estimate could be improved, we recall 
that  if a PPT state $\rho= \sum_{ijkl=0}^1|ij\>\<kl|\ot A_{ijkl}$ on 
$\ccal^{2}\ot \ccal^{2}\ot \ccal^{d_s}\ot \ccal^{d_s}$, 
%written in form $\sum_{ijkl=0}^1|ij\>\<kl|\ot A_{ijkl}$ has 
with hermitian $A_{0011}$, 
approximates a private state with shield of dimension 
(necessarily) $d_s \times d_s$,
then by \cite{KimSanders}
\be \label{KS}
\|\rho - \gamma\|_1 \geq {1\over 2(d_s +1)}.
\ee

%If we drop the assumption about hermiticity of $A_{0011}$, we obtain the following limitation  (proved in  Appendix, Sec.  \ref{sec:PPTPS}):
We will show here that the above bound holds in general for private bits,
 i.e., even if $A_{0011}$ is not hermitian.
{\proposition 
\label{thm:PPTPS}
Let  $\rho, \gamma$ be states on  $\ccal^{2}\ot \ccal^{2}\ot \ccal^{d_s}\ot\ccal^{d_s}$ 
such that $\rho$ is PPT and $\gamma$ is private. Then the bound \eqref{KS} holds. 
}

We prove this result in  Appendix \ref{sec:PPTPS}.
Thus -- in general -- the lower bound of \eqref{eq:lower} 
could not be made larger than $1 - {1\over 2(d_s + 1)}$.

Finally, one can ask if this approach could be simplified by working with separable states instead of private states. 
In Apendix  \ref{sec:pure} we will show that such approach could not work  
since the distance -- in the appropriate sense -- between a pure state and the set of PPT states 
is achieved on separable states. Consequently, one can not find a PPT state which is close to a pure state and far from separable states.

\subsection{Boosting the distance via tensoring}
%We take $k$ copies of our state $\rho$ and investigate the minimal dimension which achieves this state to be $\epsilon$ close to %private state.

We will now take $\copies$ copies of the state $\rho$ of (\ref{eq:state}) and consider 
the PPT state $\rho^{\ot \copies} $ and the private state  $\gamma^{\ot \copies}$. %denoted as $\gamma^{(\copies)}$.
By similar argument as in (\ref{eq:far}) we obtain, for any separable state $\sigma$, 
\ben
\|\rho^{\ot \copies} -\sigma\|_1&\geq& \|\sigma - \gamma^{\ot \copies}\|_1 - \|\rho^{\ot \copies} - \gamma^{\ot \copies}\|_1\nonumber \\
&\geq& \dist(\gamma^{\ot \copies},\mathcal{SEP}) - \|\rho^{\ot \copies} - \gamma^{\ot \copies}\|_1
\label{eq:derivation} \\
&\geq& 2 - {2\over 2^\copies} -\|\rho^{\ot \copies} - \gamma^{\ot \copies}\|_1 , \nonumber
\een
where the last inequality follows from 
Lemma  \ref{lem:ps_sep} and the fact that the  key-part of $\gamma^{\ot \copies}$ 
is $2^\copies\times 2^\copies$ dimensional. 
%where $\tilde{\sigma}$ is the closest separable state to $\rho^{\ot \copies}$, with locally permuted subsystems which %transformes $\gamma_1^{\ot k}$ into $\gamma^{(k)}$.
%Now, there are two inequivalent bounds on $\|\rho^{\ot k} - \gamma_1^{\ot k}\|$. 
Next, using $\|\rho^{\ot \copies} - \gamma^{\ot \copies}\|_1 \leq \copies \|\rho - \gamma\|_1$ 
(which follows by expressing $\rho^{\ot \copies} - \gamma^{\ot \copies} = \sum_{i=1}^{l} \rho^{\otimes l-i}\otimes (\rho -\gamma)\otimes\gamma^{\otimes i-1}$ \cite{Bhatia} and  by multiplicativity of $\|\cdot\|_1$  under tensoring), 
we are led to
\be
\|\rho^{\ot \copies} - \sigma\|_1 \geq 2 - {2\over 2^\copies} - {2\copies\over \sqrt{d_s} + 1}
\label{eq:first-bound}
\ee
It is now clear that by appropriately choosing $l$ and $d_s$ we can 
make the last two terms on the right as small as we wish. 
Indeed, fix $\epsilon >0$  and  let $\copies$ be the smallest integer that satisfies ${2\over 2^\copies} \leq {\epsilon\over 2}$, i.e., $\copies:=\lceil\log{ 4\over \epsilon} \rceil$.  Next, let 
$d_s$ be the smallest integer  satisfying 
${2\copies\over \sqrt{d_s} + 1}\leq {\epsilon\over 2}$, i.e., 
\be
d_s =\lceil ({4l\over \epsilon} -1)^2\rceil . 
\ee
With such choices,  we will have  $\|\rho^{\ot \copies} - \sigma\|_1 \geq 2-\epsilon$ 
for any separable state $\sigma$. Recall that  $\rho^{\ot \copies} $ is, by 
construction, a PPT state on $\ccal^{d}\ot \ccal^{d}$, where
\be
d=2^{\copies} d_s^\copies  = 2^{\copies}
\lceil ({4\copies \over \epsilon} -1)^2\rceil^{\copies} .
\ee   
Recalling that $\copies=\lceil \log \frac{4}{\epsilon} \rceil$ and streamlining 
 the formula for $d$ we obtain\begin{theorem}
\label{prop:main}
For arbitrary $\epsilon$ there exists
a 
 PPT state $\rho'$ acting  on the space 
$\ccal^{d}\ot \ccal^{d}$
with $d\leq 2^{C (\log\frac{4}{\epsilon})^2}$ and 
\be
\dist(\rho',\mathcal{SEP}) \geq 2-\epsilon.
\ee  
Here  $C>0$ is absolute contant.  The state $\rho$ is given by $\rho'=\rho^{\ot \copies}$
with $\copies=\lceil \log \frac{4}{\epsilon} \rceil$ and $\rho$ given by Eq. \eqref{eq:state}.
\end{theorem}

\noindent {\em Remark} {It is straightforward to analytically upper-bound the constant 
 $C$ by $12$;  numerically we find  $C<6$.}

%%%%%%
\smallskip 
One can obtain a slightly better estimate by appealing to  equivalence of trace distance 
and {\em fidelity} $F(\rho_1,\rho_2):= Tr\sqrt{\sqrt{\rho_2}\rho_1\sqrt{\rho_2}}$ \cite{Nielsen-Chuang} and, more precisely,   
to the (second part of the) relation  \cite{Fuchs-Graaf} 
\be \label{fidelitytrace} 
 2(1-F(\rho_1,\rho_2)) \leq \|\rho_1- \rho_2\|_1 \leq 2\sqrt{1-F(\rho_1,\rho_2)^2} 
\ee
specified to $\rho_1=\rho^{\ot \copies}$,  $\rho_2=\gamma^{\ot \copies}$. 
Since $F(\rho^{\ot \copies},\gamma^{\ot \copies}) = F(\rho,\gamma)^\copies$, 
 we can focus on calculating the fidelity between $\rho$ and $\gamma$. This is easy
since $\rho = (1-p)\gamma \oplus p \gamma'$ and so
\ben \label{fidelityrhogamma}
F(\rho,\gamma) = Tr\sqrt{\sqrt{\gamma}\rho\sqrt{\gamma}} = Tr \sqrt{(1-p) \gamma^2} = \sqrt{1-p}\ 
\een
%(The first equality is the definition of fidelity.) 
Substituting  $p = {1\over \sqrt{d_s} +1}$ and arguing as earlier we obtain
\be
\|\rho^{\ot \copies} - \sigma\|_1 \geq 2 - {2\over 2^\copies} - 2\sqrt{1- \Big({\sqrt{d_s}\over \sqrt{d_s} +1}\Big)^\copies}.
\label{eq:second-bound}
\ee
By comparing (\ref{eq:first-bound}) and (\ref{eq:second-bound}),  and then expanding in powers of 
 $\alpha=p \copies$, one finds that the above bound  is better than (\ref{eq:first-bound}).
As previously, we can deduce from  (\ref{eq:second-bound}) how the dimension 
$d$ will scale with $\epsilon$. However, the scaling is pretty much the same, possibly with a better constant.

\appendix
\section{Private states}
\label{sec:private_states}

We present here some basic properties of private states and of 
the PPT state $\rho$ of Eq. \eqref{eq:state}. 
These properties can be found, e.g., in \cite{smallkey,keyhuge}, 
while the construction itself was provided in \cite{smallkey}.
We start with the definition of private states. 
\begin{definition}
A state $\rho_{ABA'B'}$ is called a private state if it is of the form 
\be
\rho_{ABA'B'}=\sum_{i,j=1}^{d_k} {1\over d_k} |e_i\>|f_i\>\<e_j|\<f_j|\otimes U_i \sigma_{A'B'} U^\dagger_j ,
\ee
where $\{|e_i\>\}$ and $\{|f_i\>\}$ are bases in $\hcal_A$ and $\hcal_B$ respectively, $U_i$'s are unitary transformations 
acting on the system $A'B'$,  and $\sigma_{A'B'}$ is a state of that system.
\end{definition}

\noindent 
Any private state with $d_k=2$ can be written 
(up to change of basis in the key part) in the  form
\begin{eqnarray}
\gamma_{ABA'B'}=\gamma(X)={1\over 2}\left[
\begin{array}{cccc}
\sqrt{XX^{\dagger}} &0&0& X \\
0& 0&0&0 \\
0&0&0& 0\\
X^{\dagger}&0&0& \sqrt{X^{\dagger}X}\\
\end{array}
\right] ,
\label{xform}
\end{eqnarray}
where $X$ is some operator with trace norm one. 
Note that $X$ completely characterizes the private state
with $d_k=2$ (again, up to change of basis in the key part). 

We next describe the state of Eq. \eqref{eq:state} constructed in  \cite{smallkey}. 
Consider two matrices of unit trace norm:
\be
X=\frac{1}{d_s\sqrt{d_s}} \sum_{i,j=1}^{d_s} u_{ij} |ij\>\<ji|
\ee
and 
\be
Y=\sqrt{d_s} X^\Gamma = \frac{1}{d_s} \sum_{i,j=1}^{d_s} u_{ij} |ii\>\<jj| ,
\ee
where $u_{ij}$ are matrix elements of some (arbitrary) unitary 
matrix $U$ acting on $\ccal^{d_s}$ with $|u_{ij}|=1/\sqrt{d_s}$ for all $i,j$. 
For definiteness, we may set  $U$ to be quantum Fourier transform 

\be
U|k\>=\sum_{j=1}^{d_s} \sqrt{{1\over d_s}}e^{2\pi i j k  /d_s} |j\>.
\ee 
The state $\rho$ is then given by 
\be
\rho=(1-p) \gamma+ p \gamma' , 
\ee
where
\be
\label{eq:gammas}
\gamma= \gamma(X),\quad \gamma'=\sigma^x_A\ot I_{BA'B'} \gamma(Y) \sigma^x_A \ot I_{BA'B'},
\ee
with $p=\frac{1}{1+\sqrt{d_s}}$, $\sigma_x$ being a Pauli matrix.
More explicitly $\rho$ equals
\begin{eqnarray}
\frac{1}{2}\left[ \begin{array}{cccc}
(1-p)\sqrt{X X^{\dagger}} & 0 & 0 & (1-p)X \\
0 & p\sqrt{Y Y^{\dagger}}& p Y & 0 \\
0 & pY^{\dagger} & p \sqrt{Y^{\dagger}Y} & 0\\
(1-p)X^{\dagger} & 0 & 0 & (1-p)\sqrt{X^{\dagger} X}\\
%\label{eq:state2}
\end{array}
\right].
\nonumber \\
\label{eq:state2}
\end{eqnarray}

\section{Distance between PPT states and private states in finite dimension}
\label{sec:PPTPS}
For the proof of  Proposition \ref{thm:PPTPS} we need the following 
simple (and presumably well-known) lemma. 

{\lemma  \label{lem:normofgamma} 
For any operator $A$ in $\ccal^d \ot \ccal^d$ we have
\be \label{normofgamma}
\|A\|_1 \leq d\|A^{\Gamma}\|_1  \  \hbox{ and } \   \|A^{\Gamma}\|_1 \leq d\|A\|_1 .
\ee
}
The proof of the first inequality uses the following chain of (in)equalities:
\be
\|A\|_1 \leq d \|A\|_2 =d \|A^{\Gamma}\|_2\leq d \|A^{\Gamma}\|_1, \nonumber
\ee
where the equality follows from the fact that  $\Gamma$ only 
permutes the elements of a matrix, and the inequalities 
from the bounds $\|A\|_2\leq  \|A\|_1 \leq n^{1/2}  \|A\|_2$ valid for any 
$n \times n$ matrix.  The second inequality in \eqref{normofgamma}
follows then from $\Gamma$ being an involution. 

\noindent {\em Remark}: 
Note that the same bounds hold for the realignment \cite{ChenWu}, 
since it also preserves the Schatten $2$-norm.

\smallskip We now turn to the proof of 
Proposition  \ref{thm:PPTPS}. 
Let  $\rho= \rho_{ABA'B'}$ be  a PPT state and consider its block form

\be
\rho_{ABA'B'}=\left[\bea{cccc}
A_{0000} &\times&\times&A_{0011} \\
\times& A_{0101}&A_{0110}&\times \\
\times&A_{1001}&A_{1010}& \times\\
A_{1100} &\times&\times& A_{1111}  \\
 \eea
\right],
\label{blockform}
\ee
where $\times$ denotes unimportant (but not necessarily vanishing) matrix blocks. 
Our proof will be similar to that of \cite{KimSanders}. 
We assume that  $\rho_{ABA'B'}$  is 
$\epsilon$-close to some private state $\gamma$ in trace norm. 
To simplify notation, in the rest of the proof 
we will denote the trace norm $\|\cdot\|_1$ by $\|\cdot\|$.

We will use now the  so-called {\it privacy squeezing operation} 
which turns the above state into a 2-qubit state of the form 
 \be
\rho_{AB} = \left[\bea{cccc}
\|A_{0000}\| &\times&\times&\|A_{0011}\| \\
\times& \|A_{0101}\|&\|A_{0110}\|&\times \\
\times&\|A_{1001}\|&\|A_{1010}\|& \times\\
\|A_{1100}\| &\times&\times& \|A_{1111}\|  \\
\eea
\right],
\label{eq:normform}
\ee
where again $\times$ denotes unimportant but not necessarily zero matrix elements.
The operation is given by applying first unitary transformation \cite{keyhuge} of the form 
\be
\sum_{i,j=0}^1 |ij\>_{AB}\<ij| \otimes U_{ij}^{A'B'}
\ee
where $U_{00}$ and $U^\dagger_{11}$ come from the 
 singular value decomposition (SVD) of $A_{0011}$, 
and $U_{01}$ and $U^\dagger_{10}$ from the SVD of $A_{0110}$,  
and then performing partial trace over the systems $A'B'$. 
Since the state $\rho$ is PPT,  the operation applied both to the state itself, as well as to its partial transpose,
produces again a state, in particular, a positive operator. Thus 
\begin{eqnarray}
\sqrt{\|A_{0000}\|.\|A_{1111}\|} &\geq& \|A_{0011}\| , 
\label{eq:small_state}\\
\sqrt{\|A_{0101}^{\Gamma}\|.\|A_{1010}^{\Gamma}\|} &\geq&\|A_{0011}^{\Gamma}\|.
\label{eq:smallgamma}
\end{eqnarray} 
Now, since $\|\rho - \gamma\| \leq \epsilon < 1$ by assumption, 
Proposition  3 of \cite{keyhuge} implies that
\be
\|A_{0011}\|\geq {1\over 2} - \epsilon.
\ee
Hence, by \eqref{eq:small_state},  $\sqrt{\|A_{0000}\|.\|A_{1111}\|} \geq {1\over 2}-\epsilon$, 
and the  
arithmetic-geometric mean inequality shows then that $\|A_{0000}\|+ \|A_{1111}\| \geq 1-2\epsilon$. As a consequence, by the trace condition for $\rho$,
$\tr A_{0101}+ \tr A_{1010} \leq 2\epsilon$. Combining this with Eq. (\ref{eq:smallgamma}) 
and appealing again to the  arithmetic-geometric mean inequality (note that $\Gamma$ preserves the trace,
and $A_{0101}$ and $A_{1010}$ are non-negative), we obtain 
\be
\|A_{0011}^{\Gamma}\|\leq \epsilon.
\ee
In this way, we have arrived at 
\be
{\|A_{0011}\| \over \|A_{0011}^{\Gamma}\|} \geq {{1\over 2} -\epsilon \over \epsilon}.
\label{eq:last}
\ee

We now use Lemma \ref{lem:normofgamma} as it provides a bound on the left hand side of the above inequality, namely

\be
d_s \geq {\|A_{0011}\|\over \|A_{0011}^{\Gamma}\|},
\ee
which combined with (\ref{eq:last}) implies that the gap between PPT and PS states is
\be
\epsilon \geq {1\over 2(d_s + 1)}.
\ee
Thus we proved that the bound of \cite{KimSanders} 
holds in general for private bits, as asserted in Proposition  \ref{thm:PPTPS}.
  
\section{Distance between pure states and PPT states}
\label{sec:pure}
In this section we investigate the distance between a pure state and the set of  PPT states.  
It turns out that the maximal fidelity between a given pure state and  
a (arbitrary) PPT state equals 
%to 
the maximal fidelity between that pure state and a separable state. 
Consequently, as we argue below, private states
can not be replaced with pure states -- in a scheme similar to ours -- 
in order to construct a PPT state 
which is far from separable states.

{\proposition \label{prop3}
For a pure state $|\psi\>$ with Schmidt decomposition 
$|\psi\> = \sum_i a_i|e_i\>|f_i\>$ 
we have
\ben
\sup_{\sigma \in PPT} F(|\psi\>\<\psi|,\sigma)&=& \sup_{\sigma \in \mathcal{SEP}} F(|\psi\>\<\psi|,\sigma) \nonumber\\
&=&\max_i a_i =:M_a.
\label{eq:PPTfidelity}
\een
}

Before giving a proof of the  proposition let us sketch a derivation of 
its consequences mentioned earlier:  
 we can not find PPT states far from separable states by taking a pure state $|\psi\>\<\psi|\equiv \tau$ in place of private state $\gamma$ from Eq. (\ref{eq:state}). Indeed, to obtain -- for some PPT state $\rho$ -- a  bound analogous to (\ref{eq:first-bound}) via considerations going along the lines of (\ref{eq:derivation}), we would need 
  (asymptotically, when dimension is large) 
 %simultaneously  
 both 
 (i) $\dist(\tau^{\ot \copies},\mathcal{SEP})\approx 2$ and (ii) $\|\rho - \tau^{\ot \copies}\|_1 \approx 0$. The relation  \eqref{fidelitytrace} between the trace distance and fidelity would then imply (i)  $\sup_{\sigma \in \mathcal{SEP}} F(\tau^{\ot \copies},\sigma) \approx 0$ and (ii) $F(\rho,\tau^{\ot \copies}) \approx 1$. However, by the Proposition, %for any choice of $\tau$ and $\rho$ 
the  conditions (i) and (ii) can not be simultaneously satisfied since, by (\ref{eq:PPTfidelity}), the first  implies $F(\rho,\tau^{\ot \copies}) \approx 0$, which contradicts the second. 

Analogous argument shows that even the first step of the construction, Proposition \ref{almost1}, 
can not be implemented -- at least via scheme similar to ours -- with a pure state $\tau$ as a starting point. Indeed, we can not simultaneously have  $\dist(\tau,\mathcal{SEP})\geq c$ (where $c>0$ 
is a universal constant) and $\|\tau-\rho\|_1 \approx 0$. This is not entirely surprising 
since -- as is well known -- the PPT criterion perfectly discovers entanglement  
in pure states, but having precise equality of the first two quantities in (\ref{eq:PPTfidelity}) 
throughout their full range seems remarkable.

\medskip \noindent {\it Proof of Proposition \ref{prop3}} \ 
To simplify the notation, assume that $|e_i\>$ and 
$|f_i\>$ are the computational bases (the argument 
carries over {\it mutatis mutandis} to the 
general case since the set of PPT states is invariant 
under product unitary operations). Let $\sigma = 
\sum_{rstv}b_{rstv}|rs\>\<tv|$.
We want to upper-bound  $sup_{\sigma \in PPT} F(|\psi\>\<\psi|,\sigma)$. We have
\ben
F(|\psi\>\<\psi|,\sigma) &=& \sqrt{\tr (|\psi\>\<\psi|\sigma)} \nonumber \\
&= &\sqrt{\tr \biggl(\sum_{ij}a_ia_j|ii\>\<jj|\sum_{rstv}b_{rstv}|rs\>\<tv|\biggr)} \nonumber\\
&=& \sqrt{\sum_{ij}a_ia_jb_{jjii}}.
\label{eq:psi-sigma}
\een 
Given that $\sigma$ is PPT, $\sigma^{\Gamma}$ is again a state, and so 
 the inequality $b_{jjii}\bar{b}_{jjii}\leq b_{jiji} b_{ijij}$ holds for all $i,j$. Since
the elements $b_{ijij},b_{jiji}$ are diagonal, hence nonnegative, 
we can use the arithmetic-geometric mean inequality to obtain 
\be
{b_{jiji}+b_{ijij}\over 2}\geq |b_{jjii}|.
\ee
Accordingly, (\ref{eq:psi-sigma}) can be upper-bounded using the following 
chain of relations  
\ben
\sum_{ij}a_ia_jb_{jjii}& =& \text{Re} \Big(\sum_{ij}a_ia_jb_{jjii}\Big) \nonumber \\
& \leq &\Big|\sum_{i,j}a_ia_jb_{jjii}\Big|\leq \sum_{ij}a_ia_j\big|b_{jjii}\big|  \nonumber \\
&\leq& \sum_{ij}a_ia_j\Big({b_{jiji}+b_{ijij}\over 2}\Big), 
\een
where in the first equality we use the fact that fidelity is a real number, 
even though the $b_{jjii}$'s may be complex. 
Next, 
$\max_{i,j} a_ia_j \leq \max_i a_i^2 =:  M_a^2$ 
(note that $ a_i \geq 0$) and hence, 
by monotonicity of the square root function,
\ben
Tr (|\psi\>\<\psi|\sigma) &\leq& \sqrt{M_a^2 \sum_{ij}\Big({b_{jiji}+b_{ijij}\over 2}\Big)} 
 \nonumber \\
 &=&M_a =  \max_i a_i\, .
\een
This bound is easily reached by separable (in fact product) states, 
and so PPT states are as close in fidelity to 
$|\psi\>\<\psi|$ as are separable states,
which we set out to prove.

\begin{acknowledgments}
PB and KH thank Pawe\l{} Horodecki for helpful discussions.
MH acknowledges grant National Science Centre project Maestro DEC-2011/02/A/ST2/00305,
KH acknowledges Polish Ministry of Science and Higher Education
Grant no. IdP2011 000361. The research of JJ and SJS was partially supported by grants from the National Science Foundation (USA).  The contribution of JJ constituted a part of his Ph.D. thesis prepared at Case Western Reserve University. 
Part of this work has been done while KH, MH, and SJS took part
in the Quantum Information Theory program at the Institute Mittag-Leffler (Djursholm, 2010)
and when MH and SJS took part in the Mathematical Challenges 
in Quantum Information programme at the Isaac Newton Institute (Cambridge, 2013). Hospitality of 
both these institutions is gratefully acknowledged.
Part of this work has been done at the National Quantum Information Centre of Gda\'nsk.
\end{acknowledgments}

\bibliographystyle{apsrev}

%\bibliography{references3}

\end{document}